# Tunable Bandgap Opening in the Proposed Structure of Silicon Doped Graphene

Mohammad S. Sharif Azadeh[1], Alireza Kokabi[1], Mehdi Hosseini[2], Mehdi Fardmanesh[1]
*Department of Electrical Engineering, Sharif University of Technology, Tehran, Iran*
*Physics Department, Sharif University of Technology, Tehran, Iran*

Abstract- A specific structure of doped graphene with substituted silicon impurity is introduced and *ab. initio* density-functional approach is applied for energy band structure calculation of proposed structure. Using the band structure calculation for different silicon sites in the host graphene, the effect of silicon concentration and unit cell geometry on the bandgap of the proposed structure is also investigated. Chemically silicon doped graphene results in an energy gap as large as 2eV according to DFT calculations. As we will show, in contrast to previous bandgap engineering methods, such structure has significant advantages including wide gap tuning capability and its negligible dependency on lattice geometry.

Graphene, a single atomic sheet of graphite forming honeycomb lattice considered as one of the most promising crystalline forms of carbon has recently attracted significant attention [1,2,3,4]. According to its unique transport properties, remarkable electronic and structural features and electron ballistic transport, graphene is regarded as an ideal medium for many applications such as graphene-based electronic devices [5]. Two approaches are applied for fabrication of graphene: mechanical exfoliation of graphite, which is most commonly grown on Si substrate [1], and using thermal graphitization technique on SiC substrate [6,7].

Since graphene is a zero-bandgap semiconductor and exhibits semi-metallic behavior, without bandgap opening, it cannot be applied in semiconductor devices such as field-effect transistors (FETs) which cannot operate as expected in the absence of bandgap in the material [1]. On the other hand, creating tunable bandgap may ultimately enable enormous applications of graphene in digital electronics, pseudo-spintronics [8], terahertz technology [9], and infrared nano-photonics [10]. Thus, the graphene community is greatly motivated to the tantalizing research topic of band splitting.

In the quest to opening an energy gap, several physical and chemical approaches such as creating local strain, use of graphene nanoribbons (GNR), periodic hydrogenation (graphane), graphene-substrate interaction and patterning graphene are proposed or implemented. Han *et al.* observed a tunable bandgap for thin strips of graphene, which is a function of GNR's width [11]. However, currently there is no helpful method to implement nanoribbons with desirable nanometer scale width. Moreover, nanoribbon devices commonly have low driving currents [12].

Epitaxially grown graphene on the SiC substrate is supposed to have strong coupling with the substrate and results in the elimination of the A and B sublattices symmetry. Following this idea, Zhao *et al.* predicted a gap of about 260meV in Dirac points for the epitaxially grown graphene on SiC substrate due to graphene-substrate interaction [13]. However, this gap decreases as the sample thickness increases in the multi layer structures and totally disappears when the number of layers increases only to four [14].

Since any kind of defects in symmetrical graphene layer might cause bandgap opening, there has been tremendous interest in the application of patterning methods in this field. Proposing a specific patterned structure of the graphene, Tiwari *et al* [15] have reported an electronic bandgap as large as 65meV for the gate voltages less than 1.5V. Based on a similar principle, vacancy clusters called antidot lattice was another approach to achieve gapped graphene [16]. Such a patterned graphene is developed simultaneously in two different groups most recently [17,18].

Deep-level impurities, displaying many intricate transport properties is a well-known bandgap engineering method in the semiconductor technology applied to adjust the electric properties of the host material using in-crystal-site doping or combination stoichiometry [19]. Inspired by this method, here, we wish to draw attention to the possibility of forming silicon deep-level doping in the graphene. We discuss the formation of a graphene-based honeycomb lattice in which some of the crystal sites are semi-randomly substituted by their chemically analogous silicon atoms. In contrast to antidot lattice, in the discussed structure, called here siliphene for simplicity, the vacancies of the host crystal are considered to be passivated by the silicon doping as impurities.

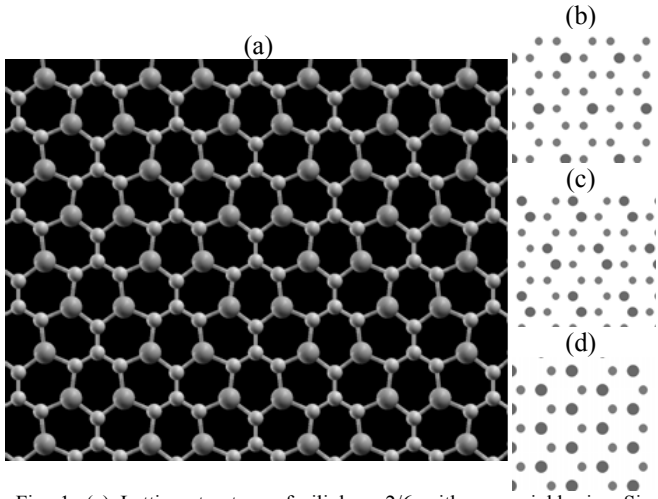

Fig. 1. (a) Lattice structure of siliphene:2/6 with non-neighboring Si atoms. This structure is obtained from force relaxation. In other structures, simplified lattices of (b) siliphene:1/6, (c) siliphene:2/6 with neighboring Si atoms and (d) siliphene:1/2 are shown. In the proposed numbering method, the first number denotes the number of silicon atoms.

According to the extreme progress on the fabrication of the graphene-based structures [20,21], the possibility of implementation of siliphene will be discussed later in this paper.

The lattice of semi-randomly distributed silicon atoms is hoped to have the important consequence that it turns the uniform and symmetric semi-metallic graphene into a gapped periodic structure, which is supposed to show semiconducting behavior. Similar methods are previously applied to heterostructure semiconductors for band structure engineering.

By a similar principle, silicon to carbon stoichiometry, $Si_xC_{1-x}$ or silicon doping level adjusts the band structure of the siliphene.

Assuming siliphene to be a graphene-based semiconductor, it would be an ideal medium for the FET fabrication due to high electron mobility of graphene, which is possibly inherited by siliphene. Such high electron mobility can lead to fabrication of very high switching frequency transistors.

Here, we predict the formation of tunable electronic energy gap in the siliphene structure using density-functional theory (DFT) calculations in which the many-body problem of interacting electrons in the static potential is reduced to the problem of non-interacting electrons moving in an effective potential. Since the covalence bonds in the siliphene involve electrons of $s$ and $p$ orbitals, it is expected that mean-field DFT approach should be valid in the calculation of electronic band structure.

Electronic structure calculations have been performed using *ab initio* full potential linear augmented plane waves (FP-LAPW). We have used the Wien2k computational code [22], which is a powerful code in this area. The initial Si-C and C-C bond length parameters for the siliphene have been taken from experimental data as 1.91Å and 1.42Å, respectively [23]. Then, the equilibrium lattice parameters have been obtained from force relaxation process. The radii of muffin-tin spheres of C and Si ions have been considered to be approximately 1.3 and 1.65a.u., respectively, and have been chosen such that they remain non-overlapping in the volume optimization and electronic structure calculation. We have used the generalized gradient approximation (GGA) with Perdew–Burke–Ernzerhof parameterization in determining the exchange and correlation energies [24]. The energy to separate core and valence states has been set as a value of −8Ry. The RMT×$K_{max}$ and cut off angular momentum $L_{max}$ parameters have been considered for these calculations as 7.0 and 10, respectively. The summations have been carried out on 800 k-points corresponding to 48k points in the irreducible wedge of the Brillouin zone. This number of k points is in correspondence with a 9×9×7 mesh in the Monkhorst–Pack scheme.

In the proposed structures, the basis consists of a hexagon with six atoms of its corners, which at least three of them are carbon atoms and the remaining are silicon ones. The crystalline structure is a triangular lattice of the considered basis. All of the considered structures are depicted in Fig 1. The only distinct structure of two non-neighboring Si atoms in a honeycomb unit cell is shown in Fig. 1a. As illustrated in Fig. 1b, siliphene:1/6, only one of six corners of the hexagon is occupied by Si atom. In Fig. 1c, the third structure, siliphene:2/6 with two neighboring Si atoms is demonstrated. Finally, Fig. 1d is for demonstration of siliphene:1/2 structure which contains a bonded carbon and silicon atoms in its unit cell.

It should be noted that the stability of all of the depicted structures are verified using the DFT calculations in two dimensions. Making considered structures relaxed by QUANTUM ESPRESSO, leads to deformation of the lattice constants with respect to graphene ones while the new structure remains 2D.

The anisotropic two-dimensional effective mass can be expressed by [25]

$$m^{*-1}_{c,v_{x,y}} = \hbar^{-2}\left|\frac{\partial E_{c,v}}{\partial \kappa^2_{x,y}}\right| \quad (1)$$

For the graphene, in Dirac points the effective mass is equal to zero. Therefore, the carrier mobility in graphene theoretically diverges. Carrier mobility in graphene is reported experimentally to be as high as 100000 [26]. However, the experimentally achieved mobility is lower than expected value due to effect of periodic potential of the substrate, scattering by the impurities and electron-phonon interactions. Usually bandgap opening leads to increase of effective mass and decrease of mobility because of defect scattering. For instance, antidot [16] and nanomesh [17] cause electrons to be scattered by vacancies. It can be predicted that silicon doping in graphene structure, has less destructive effect on graphene mobility due to resemblance of Si and C atoms. The effective mass of electron and hole carriers are calculated by DFT calculations and shown in Table 1. Certainly, the effective masses comparing other bandgap opening methods had a little negative effect on carrier effective masses and higher mobility is predicted.

| Number of Si in unit cell | Energy Gap(eV) | Effective mass ||
|---|---|---|---|
| | | $m_c^*/m_0$ | $m_v^*/m_0$ |
| 0[a] | 0 | - | - |
| 1/6[b] | 0.1308 | 0.2716 | 0.1992 |
| 3/12 | 0.8375 | 0.1849 | 0.1823 |
| 2/6[c] | 1.2372 | 0.2119 | 0.2412 |
| 2/6[d] | 1.5714 | 0.2018 | 0.2183 |
| 1/2 | 2.0161 | 0.1528 | 0.1325 |
| 2/2[e] | 0 | - | - |

Table 1. The energy gap and effective mass for selected structures of siliphene.
[a] Graphene structure
[b] One of six atoms of unit cell is replaced by Si
[c] Two replaced Si atoms are in neighborhood
[d] Two replaced Si atoms are non-neighboring

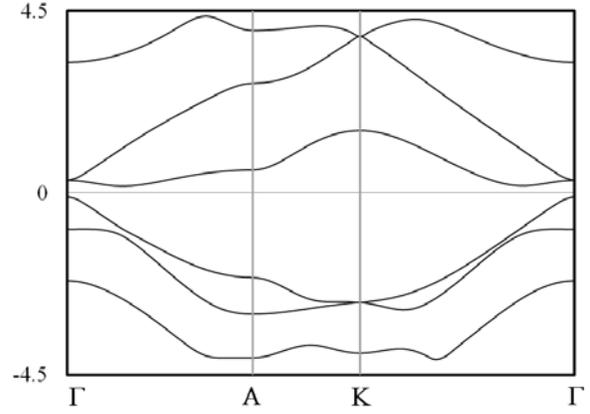

Fig. 3. The band structure of siliphene:1/6 structure

As it is obvious from the table, the lowest effective mass is obtained for the 1/2 structure, which is also the most symmetrical structure among the silicon-doped ones. This result is also intuitively expected for the carrier effective mass in the periodic structure of the crystal. The gap induced in siliphene by doping silicon in the graphene is controllable with the amount of silicon content. Since the largest direct gap is also acquired from this structure, it might be a promising choice for the semiconducting applications. In the first row of the table, the data of the graphene is shown, which has no electronic gap indeed. From the second row, the electronic bandgap and the carrier effective mass of the structures with further silicon content are depicted sequentially. It is remarkable that the calculated gaps are all direct ones except in the siliphene:1/6 structure which indirect gap is observed.

As illustrated is Fig. 2, siliphene:1/2 structure has a direct energy gap of 2eV in $k$ point. Since the top of valance band is under Fermi energy level, this material has semi-conducting behavior. Fig. 3 shows the $E$-$k$ diagram for siliphene:1/6 structure. As the top of valance band and the bottom of conductance band are not aligned, this structure leads to an indirect band gap, unlike the other considered siliphene structures. These results are acknowledged by density of function diagram.

The carrier density in the graphene is in the order of $10^{12} cm^{-2}$ [27]. Here we use a simple model for the calculation of the carrier density in the proposed structure based on the numerical results obtained from the DFT analyses. The expression for the carrier density in an intrinsic semiconductor can be written as [25]

$$n_c(T) = \int_{E_c}^{\infty} g(E) f_e(T) dE$$
$$p_v(T) = \int_{-\infty}^{E_v} g(E) f_h(T) dE \quad (2)$$

where $g(E)$ is the density of states in semiconductor, which is calculated by DFT method using Wien2k code and shown in Fig. 2.b, and $f_e(T)$ and $f_h(T)$ are Fermi-Dirac distribution for the electrons and holes, respectively. The carrier densities for the electrons ($n_i$) and holes ($p_i$) in siliphene:1/2 structure at $T$=300K are obtained numerically from the equation (2) to be $7.13 \times 10^3 cm^{-2}$ and $8.73 \times 10^3 cm^{-2}$, respectively. In this case, we approximated the density of the states at $T$=300K by the ones at $T$=0K resulted from DFT method. Therefore, as expected, the carrier densities for electrons and holes in siliphene are approximately equal since the Fermi level is in mid-gap position. Thus, siliphene:1/2 behaves as an intrinsic semiconductor, which can be doped by *p* or *n* dopants. However, siliphene:1/6, in which the Fermi level is lower than the middle of the gap, behaves as moderate *p*-type semiconductor. Here, $n_c$ is calculated to be $1.42 \times 10^{11}$ where $p_v$ is found to be $7.03 \times 10^{11}$ at 300K. Thus, the type of the semiconductor is also affected by silicon doping. The gained energy bandgap can be tuned by Si dopant ratio, rather than dopants geometry, since two different structures of siliphene:2/6 has near energy gaps, as illustrated in Fig. 4. The interpolating line shows that doping more silicon would result in formation of a larger bandgap and this trend is established for Si dopant level less than ½ at least for the considered structures. Silicene, theoretical equivalent of graphene for silicon, is supposed to be gapless [28]. Assuming the silicene as the ending point of silicon content addition, and a continuous behavior, one would expect the downturn of the mentioned trend beyond a specific point.

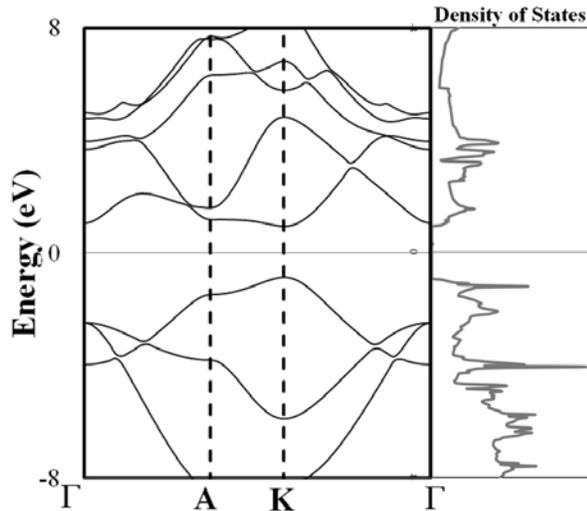

Fig. 2. The band structure and associated density of states for siliphene:1/2 structure. The Fermi level is in mid-gap position.

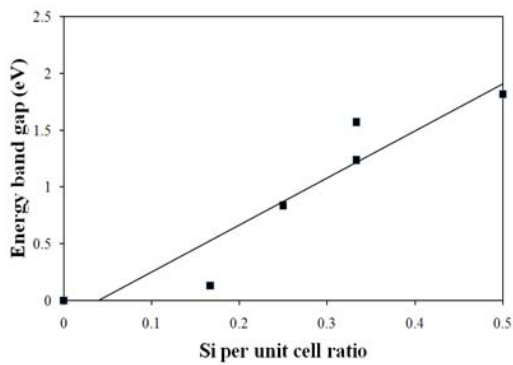

Fig. 4. The calculated energy band gaps versus the number of Si atoms in siliphene unit cell: 0, 1/6, 3/12, 2/6 and 1/2.

As mentioned, the DFT analysis predicts that the silicon-doped graphene could be stable in a 2D structure due to atomic force relaxation calculations. This result is theoretically in favor of the possibility of forming such structure in advance. As it is obvious, moving from graphene to siliphene structure requires some of the carbon atoms to be replaced by silicon ones. In upcoming future, this movement might become possible as many modern technologies are being introduced and applied in the fabrication of graphene and other nanostructures. Among these technologies, some of them seem to be more applicable regarding to their recent applications in the relevant empirical works. Such technologies are molecular beam epitaxy (MBE), which is widely applied in the fabulous heterostructures implementation, reactive ion-beam etching being used in nanometer patterning, Micromechanicall cleavage of SiC and state-of-the-art e-beam lithography employed to carve graphene nanoribbons. Especially, MBE seems to be more promising following the very recent progress in the fabrication of single and bi-layer graphene using this method.

In summary, we proved that the theoretically subjected silicon-doped graphene sheet shows semiconducting behavior. The bandgap has dependency on Si per unit cell ratio. For the ratios lower than 0.5, the more Si atoms in siliphene unit cell the higher bandgap is obtained. The observed bandgap in Dirac points is direct and its value can be tuned by dopant proportion in unit cell. The obtained bandgap has a minor dependency of Si atoms locations in unit cell, but it is found to be mainly a function of Si per unit ratio. Calculated effective masses for carriers in siliphene are lower than commonly used semiconductors such as Si and GaAs and can lead to higher mobility and higher switching frequency. We also showed that this proposed structure is stable in two-dimensions. Next steps are p-n doping of siliphene by the atoms of III and V group such as Baron and Phosphorus atoms and proposing semiconductor devices using this graphene based structure.